\documentclass{aa}

\usepackage{graphicx}
\usepackage{txfonts}

\begin{document}
\title{Separated before birth: pulsars B2020+28 and B2021+51
as the remnants of runaway stars}

\author{V.V.Gvaramadze\inst{1}}

\institute{Sternberg Astronomical Institute, Moscow State
University, Universitetskij Pr. 13, Moscow 119992, Russia;\\
vgvaram@sai.msu.ru}

\date{Received 24 April 2007/ Accepted 22 May 2007}

\titlerunning{Pulsars B2020+28 and B2021+51}
\authorrunning{Gvaramadze}

\abstract{Astrometric data on the pulsars B2020+28 and B2021+51
suggest that they originated within several parsecs of each other in
the direction of the Cyg OB2 association. It was proposed that the
pulsars share their origin in a common massive binary and were
separated at the birth of the second pulsar following the asymmetric
supernova explosion. We consider a different scenario for the origin
of the pulsar pair based on a possibility that the pulsars were
separated before their birth and that they are the remnants of
runaway stars ejected (with velocities similar to those of the
pulsars) from the core of Cyg OB2 due to strong three- or four-body
dynamical encounters. Our scenario does not require any asymmetry in
supernova explosions.

           }

\maketitle

\section{Introduction}
%
Most stars form in dense embedded clusters and reside in binary
systems (either primordial or tidal). If both binary components are
massive enough, they end their lives as core-collapsed supernovae
(SNe). Stellar remnants of SN explosions, usually the neutron stars
(NSs), have peculiar velocities at least an order of magnitude
higher than the typical velocities of their progenitors, the OB
stars (e.g. Gunn \& Ostriker \cite{gun70}). It is believed that the
high velocities of NSs are due either to the asymmetry of SN
explosions (e.g. Dewey \& Cordes \cite{dew87}) or to the disruption
of tight massive binaries following the second (symmetric) SN
explosion (e.g. Iben \& Tutukov \cite{ibe96}). The progress in
measuring the proper motions and parallaxes of NSs (pulsars) allows
their peculiar (transverse) velocities to be determined with high
precision and makes it possible to trace their trajectories back to
the parent star clusters (e.g. Hoogerwerf et al. \cite{hoo01};
Chatterjee et al. \cite{cha05}). Recently Vlemmings et al.
(\cite{vle04}; hereafter VCC04) have used the high-precision
astrometric data (proper motions and parallaxes) for two dozen
pulsars to determine their trajectories in the Galactic potential
and to search for pairs with a common origin. They discovered that
two pulsars from their sample (presently separated by $\sim
23^{\degr}$) originated within several parsecs of each other in the
direction of the \object{Cyg OB2} association. VCC04 interpret their
discovery as an indication that the progenitors of both pulsars,
\object{B2020+28} and \object{B2021+51}, were the members of a
common massive binary and suggest that the pulsars were separated at
the birth of the second one following the asymmetric SN explosion.

In this Letter we explore a different scenario for the origin of
B2020+28 and B2021+51. We suggest that these pulsars were separated
before their birth and that they are the remnants of runaway stars
ejected (with velocities similar to those of the pulsars) from the
parent star cluster due to the strong three- or four-body dynamical
encounters. Our scenario does not require any asymmetry in SN
explosions.

\section{Pulsars B2020+28 and B2021+51: origin in a common binary}

The main result presented in VCC04 is that B2020+28 and B2021+51
originated within several parsecs of each other. VCC04 derived the
most likely three-dimensional peculiar velocities of the pulsars at
birth, $\simeq 150$ and $\simeq 500 \, {\rm km} \, {\rm s}^{-1}$
(respectively, for B2021+51 and B2020+28), and the angle between the
velocity vectors $\psi \simeq 160^{\degr}$. These velocities can, in
principle, be produced via disintegration of a tight (semi-detached)
massive binary after the second (symmetric) SN explosion (e.g. Iben
\& Tutukov \cite{ibe96}); in this case, however, $\psi$ is always $<
90^{\degr}$. Moreover, the spin characteristics of B2020+28 and
B2021+51 (typical of non-recycled pulsars) argue against the origin
of these pulsars in a common {\it tight} binary (cf. VCC04).

One possible way to reconcile the kinematic data with the common
binary scenario is to assume that the binary was disrupted either
after the first or the second {\it asymmetric} SN explosion (VCC04).
Note that the similarity between the pulsar's characteristic ages
($\simeq 2.88$ and $\simeq 2.75$ Myr) implies that the mass ratio of
the binary components was $\sim 1$. Therefore, depending on the
initial parameters (binary separation, etc), the binary system at
the moment of the first SN explosion consists of two red supergiant
or Wolf-Rayet stars or of two carbon-oxygen (CO) cores.

The latter situation can be realised if the massive binary evolves
through two common-envelope phases (see Belczy\'{n}ski \& Kalogera
\cite{bel01}). A natural outcome of the evolution of this type of
binaries, provided that the SN explosions were of zero or moderate
asymmetry, is the origin of a binary {\it non-recycled} pulsar
(Belczy\'{n}ski \& Kalogera \cite{bel01}). The CO binary, however,
could be disrupted after the first (or the second) asymmetric SN
explosion if the kick received by the stellar remnant was of proper
magnitude and orientation (see Tauris \& Takens \cite{tau98}).

For illustrative purposes, we consider the disruption of a CO binary
following the first asymmetric SN explosion. For parameters of the
CO binary given in Belczy\'{n}ski \& Kalogera (\cite{bel01}) and
using Eqs.\,(44)--(47) and (51)--(56) given in Tauris \& Takens
(\cite{tau98}), one can show that the pulsar velocities and $\psi$
could be explained if the kick imparted to the first-born pulsar
(B2020+28)
\begin{figure}
 \resizebox{8cm}{!}{\includegraphics{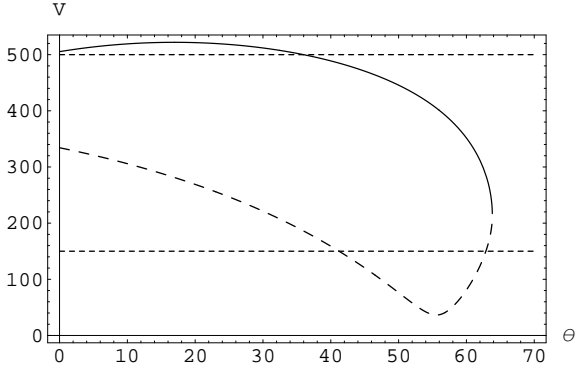}}
 \caption{The dependence of the velocities of the (first-born)
pulsar and its former companion star (now the runaway
progenitor of the second pulsar) on the angle between the kick vector and
the direction of motion of the exploding star (shown, respectively, by the
solid and the long-dashed lines). The horizontal short-dashed lines indicate
the pulsar velocities suggested by VCC04. See text for details.}
  \label{twovel}
\end{figure}
was $\sim 500 \, {\rm km} \, {\rm s}^{-1}$ (for the sake of
simplicity we assume that the second SN explosion was symmetric),
while the angle between the kick vector and the direction of motion
of the exploding star, $\theta$, was $\simeq
40^{\degr}$\footnote{Note that for $64^{\degr} \la \theta \la
290^{\degr}$, the binary system remains bound.} (see Figs.\,1 and 2
and Gvaramadze \cite{gva06}). It is obvious that the kick should be
stronger if, at the moment of the first SN explosion, the binary
consists of red supergiant or Wolf-Rayet stars (cf. VCC04).

Another possibility is that the pulsars attained their velocities in
the course of disintegration of the binary after the second
asymmetric SN explosion. Since both pulsars are not recycled, one
should assume either that the binary separation was sufficiently
large (so that the wind of the secondary star did not affect the
evolution of the first-born pulsar) or that the binary evolved
through a double common-envelope phase (see above). VCC04 suggest
that the pulsars were born in a wide binary, but in their analysis
they draw an erroneous conclusion that the pulsar velocities can be
explained by a kick of only $\simeq 200 \, {\rm km} \, {\rm s}^{-1}$
(see Gvaramadze \cite{gva06}). One can show, however, that in both
the above-mentioned cases the kick imparted by the second SN
explosion should be $\geq 500 \, {\rm km} \, {\rm s}^{-1}$.
\begin{figure}
 \resizebox{8cm}{!}{\includegraphics{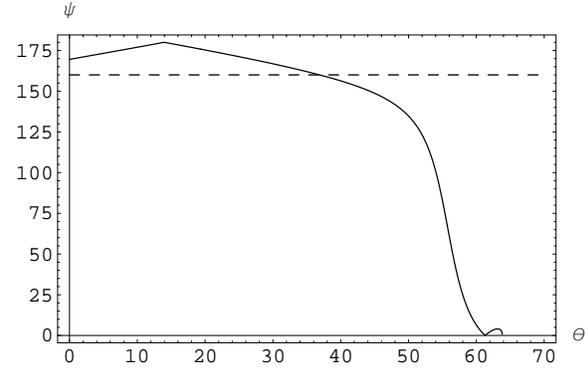}}
 \caption{The angle between the velocity vectors of the first- and
second-born pulsars as a function of the angle between the kick
vector and the direction of motion of the exploding star. The horizontal
dashed line indicates the angle between the pulsar velocity vectors
suggested by VCC04.}
  \label{chitheta}
\end{figure}

Thus, the origin of the pulsars in a common binary implies that at
least one of the SN explosions was asymmetric enough to produce a
kick of $\geq 500 \, {\rm km} \, {\rm s}^{-1}$. If, however, SNe can
indeed impart high velocities to NSs, then it is not necessary to
assume that the pulsars originated in the same binary, but instead
one can suggest that they were created by two separate SN explosions
occurred in the same parent star cluster within a few $10^5$ yr. Our
scenario for the origin of the pulsar pair has something in common
with the latter possibility, but we do not require any asymmetry in
the SN explosions.

\section{Pulsars B2020+28 and B2021+51: dynamical ejection from the young
massive star cluster}

The recent discovery of the so-called hypervelocity stars (Brown et
al. \cite{bro05}) and hyperfast pulsars (Chatterjee et al.
\cite{cha05}), the ordinary stars and pulsars moving with extremely
high ($\sim 1\,000 \, {\rm km} \, {\rm s}^{-1}$) peculiar
velocities, suggests a possibility that the hypervelocity stars
could be the progenitors of hyperfast NSs, provided that they are
massive enough (Gvaramadze et al. \cite{gva07}). A strong argument
in support of this possibility comes from the fact that the mass of
one of the hypervelocity stars, HE\,0437$-$5439, is $\ga 8 \,
M_{\odot}$ (Edelmann et al. \cite{ede05}) so that, in principle, it
can end its evolution as a hyperfast NS! The high velocities ($\sim
200-400 \, {\rm km} \, {\rm s}^{-1}$) inferred for some early B-type
stars at high galactic latitudes (Ramspeck et al. \cite{ram01}) also
support the possibility that high-velocity pulsars could originate
from high-velocity runaway stars.

Gvaramadze et al. (\cite{gva07}) suggest that the origin of
hypervelocity stars could be connected not only with scattering
processes involving the supermassive black hole (BH) in the Galactic
centre (the common wisdom; originally suggested by Hills
\cite{hil88}; see also Yu \& Tremaine \cite{yu03}; Gualandris et al.
\cite{gua05}), but also with strong three- or four-body dynamical
encounters in the dense cores of young massive star clusters
(YMSCs), located either in the Galactic disk or near the Galactic
centre. The discovery of a halo population of early B stars, whose
lifetimes are shorter than the times-of-flight from the Galactic
centre (Brown et al. \cite{bro07}; see also Ramspeck et al.
\cite{ram01}), supports this suggestion. We believe, therefore, that
the pulsars B2020+28 and B2021+51 could be the remnants of
high-velocity runaway stars ejected from the same YMSC. The
kinematic and characteristic ages of the pulsars (respectively,
$\sim 2$ and $\sim 3$ Myr; VCC04) imply that by the moment of
ejection the progenitor stars have already become Wolf-Rayet stars
[the short-lived ($< 1$ Myr) helium (He) cores of massive stars; cf.
Gvaramadze et al. \cite{gva07}].

Of the two mechanisms that could be responsible for the origin of
the high-velocity progenitors of B2020+28 and B2021+51, the first
relies on close dynamical encounters between hard (Heggie
\cite{heg75}) massive binary stars in the dense core of a YMSC. The
peculiar velocities of runaway stars produced in this process are
similar to the orbital velocities of the binary components (e.g.
Leonard \& Duncan \cite{leo90}), but occasionally they could be much
higher. Scattering experiments by Leonard (\cite{leo91}) showed that
the maximum velocity attained by the lightest member of the binaries
involved in the interaction (e.g. the He core of a massive star or
an early B-type star) can be as high as the escape velocity, $V_{\rm
esc}$, from the surface of the most massive star in the binaries.

For the main-sequence stars with the mass-radius relationship
(Habets \& Heintze \cite{hab81}), $r_{\rm MS} = 0.8 (m_{\rm MS}
/M_{\odot} )^{0.7} \, R_{\odot}$, where $r_{\rm MS}$ and $m_{\rm
MS}$ are the stellar radius and the mass, the maximum possible
velocity of ejected stars is a weak function of $m_{\rm MS}$,
$V_{\rm ej} ^{\rm max} \simeq V_{\rm esc} \simeq 700 \, {\rm km} \,
{\rm s}^{-1} (m_{\rm MS} /M_{\odot} )^{0.15}$ and could be as high
as $\sim 1\,400 \, {\rm km} \, {\rm s}^{-1}$ (cf. Leonard 1991).
Numerical simulations by Leonard (\cite{leo91}) showed that about $4
\%$ of binary-binary encounters result in the origin of runaway
stars with $V_{\rm ej} \simeq 0.5V_{\rm esc}$, which is enough to
explain the velocity of $\sim 500 \, {\rm km} \, {\rm s}^{-1}$
suggested by VCC04 for one of the pulsars. Note that the results of
Leonard (\cite{leo91}) were used by Tenjes et al. (\cite{ten01}) to
explain the origin of the high-velocity ($\sim 400 \, {\rm km} \,
{\rm s}^{-1}$) runaway star HIP 60350.

Another possible mechanism for producing high-velocity stars is
based on exchange encounters between tight binary stars and a
compact massive object, either a very massive star (VMS), formed
through the runaway stellar collisions and mergers in the mass
segregated core of a YMSC (e.g. Portegies Zwart et al.
\cite{por99}), or its descendant, an intermediate-mass BH (e.g.
Portegies Zwart \& McMillan \cite{por02}). After the close encounter
and tidal breakup of the binary, one of the binary components
(usually the more massive one) becomes bound to the compact object,
while the second one recoils with a high velocity given by $V_{\rm
ej} \sim [M/(m_1 +m_2)]^{1/6} (2Gm_1 /a)^{1/2}$ (Hills \cite{hil88};
see also Gvaramadze et al. \cite{gva07}), where $M$ is the mass of
the compact object, $m_1$ and $m_2$ are the masses of the binary
components ($m_1 >m_2$), and $a$ the binary separation.

In YMSCs of mass $\sim 10^4 \, M_{\odot}$, the mass of the VMS does
not exceed several $100 \, M_{\odot}$, while the thermal
(Kelvin-Helmholtz) time scale of the VMS is shorter than the mean
time between collisions (see Portegies Zwart et al. \cite{por99}).
In this case, the growing VMS rapidly evolves to the thermal
equilibrium (e.g. Suzuki et al. \cite{suz07}), so that one can adopt
the following mass-radius relationship, $R \simeq 1.6 \,
(M/M_{\odot} )^{0.47} R_{\odot}$, where $R$ is the radius of the VMS
(see Freitag et al. \cite{fre06} and references therein). In the
process of an exchange encounter with a binary, the VMS could be
considered as a point mass if the binary tidal radius, $r_{\rm t}
\sim [M/(m_1 +m_2)]^{1/3} a$, is at least several times larger than
$R$. For $M=200-300 \, M_{\odot}, m_1 =30 \, M_{\odot}$ (a
main-sequence star), $m_2 =8\, M_{\odot}$ (a He core), and $a=50 \,
R_{\odot}$, one has $r_{\rm t} \simeq 90-100 \, R_{\odot}$ (i.e.
much larger than $R\simeq 19-23 \, R_{\odot}$) and $V_{\rm ej}
\simeq 630-670 \, {\rm km} \, {\rm s}^{-1}$, that is enough to
explain the pulsar velocities.

In more massive ($\geq 10^5 \, M_{\odot}$) YMSCs, the VMS can
acquire a mass of several $1\,000 \, M_{\odot}$ (Portegies Zwart et
al. \cite{por04}). But in this case, the thermal time scale is
comparable to the collision time and the VMS remains inflated untill
collapsing into an intermediate-mass BH (e.g. Portegies Zwart et al.
\cite{por06}). Exchange encounters with this VMS would not produce
high ejection velocities. The star ejection from the YMSC, however,
would be very effective if the VMS leave behind a BH of mass $\sim
1\,000 \, M_{\odot}$ (e.g. Gualandris \& Portegies Zwart
\cite{gua07}).

\section{Cyg OB2}

The astrometric data on B2020+28 and B2021+51 suggest that these
pulsars (or their progenitors; our preference) were ejected $\sim
1.9$ Myr ago from the same origin at a distance of $\sim 1.9$ kpc in
the direction of the Cyg OB2 association (VCC04). The parent YMSC
(or its descendant) should still be located at about the same
distance since its possible peculiar velocity of $\leq 30 \, {\rm
km} \, {\rm s}^{-1}$ (typical of the OB associations near the Sun;
de Zeeuw et al. \cite{dez99}) would result only in a slight offset
of $\leq 60$ pc (cf. VCC04). To constrain the current age of the
parent cluster, we assume that the initial mass of the progenitor
stars of B2020+28 and B2021+51 could be as high as $\ga 50 \,
M_{\odot}$. (It is believed that stars of this mass can lose most of
their mass via stellar wind or mass transfer on a binary companion
and leave behind NSs; e.g. Vanbeveren et al. \cite{van98}; Wellstein
\& Langer \cite{wel99}; cf. Muno et al. \cite{mun06}.) From this it
follows that the minimum age of the parent YMSC should be $\sim 5$
Myr, that is, $\sim 2$ Myr (the pulsar kinematic age) plus $\sim 3$
Myr (the lifetime of a $\ga 50 \, M_{\odot}$ star). Assuming that
the YMSC initially contained at least 10 stars of mass $> 50
M_{\odot}$, one has the (initial) mass of the cluster of $\geq 10^4
\, M_{\odot}$ (for a $0.2-120 \, M_{\odot}$ Salpeter initial mass
function).

The only likely candidate for the birth cluster of B2020+28 and
B2021+51 in the region suggested by VCC04 is the Cyg OB2
association. Numerous star clusters in its neighbourhood (see, e.g.,
Le Duigou \& Kn\"{o}dlseder \cite{led02}) cannot pretend to play
this role either due to their youth or low masses.

Cyg OB2 is one of the most massive and compact OB associations in
our Galaxy (Kn\"{o}dlseder 2000). The large number ($\sim 100$) of O
stars identified in Cyg OB2 (Kn\"{o}dlseder \cite{kno00}; see also
Comer\'{o}n et al. \cite{com02}) implies that its mass could be as
high as $\sim 10^5 \, M_{\odot}$. The angular radius of Cyg OB2 is
$\sim 1^{\degr}$, while the half light radius is $\sim 13^{'}$
(Kn\"{o}dlseder \cite{kno00}), that at the distance of Cyg OB2 of
$\sim 1.5-1.7$ kpc (Hanson \cite{han03}; Massey \& Thompson
\cite{mas91}) corresponds, respectively, to $\sim 25-30$ pc and
$\sim 5-6$ pc. Note that the centre of Cyg OB2 lies within the
$2\sigma$ likelihood contour of the pulsar birth location and, at
the $2\sigma$ level, the distances to the Cyg OB2 and the birth
location are consistent with each other. Age estimates for Cyg OB2
range from $\sim 1$ to 5 Myr (e.g. Bochkarev \& Sitnik \cite{boc85};
Kn\"{o}dlseder et al. \cite{kno02}). The wide age spread suggests
that the star formation in the Cyg OB2 was non-coeval. The
non-coevality could be understood if the star formation in the
association started initially in the dense core of the parent
molecular cloud and then propagated to its lower density periphery.
It is believed (e.g. Elmegreen \cite{elm00}) that the star formation
occurs on one or two dynamical time scales, $t_{\rm dyn} \sim (G\rho
)^{-1/2}$, where $\rho$ is the gas density in the cloud, so that in
a density-stratified cloud of mass of $\sim 10$ times higher than
the stellar mass of Cyg OB2 and the size similar to that of the
association, the age spread could be comparable with $t_{\rm dyn}
\sim 5$ Myr.

If the progenitor stars of B2020+28 and B2021+51 were ejected from
Cyg OB2, then we suppose that, $\sim 2$ Myr ago (or $\sim 3$ Myr
after the formation of the first massive stars in the centre of the
association), the core of the association was much more compact and
denser. Assuming that the association expands with a velocity equal
to its velocity dispersion ($\simeq 2.4 \, {\rm km} \, {\rm
s}^{-1}$; Kiminki et al. \cite{kim06}), one finds that the stars
located within the current half light radius were originally
concentrated in a region of a radius of $< 1$ pc. It is likely that
the two star clusters projected close to each other ($\sim 6^{'}$)
near the centre of Cyg OB2 (Bica et al. \cite{bic03}) are the
remainders of this dense core. We suggest that the mass of the core
was much higher ($\geq 10^4 \, M_{\odot}$) than the current mass of
the clusters (several $1\,000 M_{\odot}$; Bica et al. \cite{bic03})
and that it was significantly reduced during the last 2 Myr due to
the overall expansion of the association and star ejections
following close dynamical encounters [recent N-body simulations by
Pflamm-Altenburg \& Kroupa (\cite{pfl06}) showed that dynamical
processes in the Trapezium cluster could be responsible for the loss
of at least $75\%$ of its initial content of OB stars]. Thus we
believe that $\sim 2$ Myr ago the conditions in the core of Cyg OB2
were favourable to the dynamical processes discussed in Sect.\,3.

\begin{acknowledgements}
I am grateful to H.Baumgardt, D.Bomans, P.Kroupa, and S.Portegies
Zwart for useful discussions. I am also grateful to D.Bomans and
R.-J.Dettmar for their hospitality during my stay at the
Astronomisches Institut, Ruhr-Universit\"{a}t Bochum, where this
work was partially carried out. This work was partially supported by
the Deutsche Forschungsgemeinschaft.
\end{acknowledgements}

\end{document}